
%
%
%
%
%
%
%
%
%
%
%
\nonstopmode
\catcode`\@=11 
%
%
%
\font\seventeenrm=cmr17

\font\twelverm=cmr12
\font\ninerm=cmr9
\font\sixrm=cmr6

\font\seventeenbf=cmbx12 at 17pt
\font\fourteenbf=cmbx12 at 14pt
\font\twelvebf=cmbx12
\font\ninebf=cmbx9
\font\sixbf=cmbx6

\font\seventeeni=cmmi12 at 17pt             \skewchar\seventeeni='177
\font\fourteeni=cmmi12 at 14pt              \skewchar\fourteeni='177
\font\twelvei=cmmi12                        \skewchar\twelvei='177
\font\ninei=cmmi9                           \skewchar\ninei='177
\font\sixi=cmmi6                            \skewchar\sixi='177

\font\seventeensy=cmsy10 scaled\magstep3    \skewchar\seventeensy='60
\font\fourteensy=cmsy10 scaled\magstep2     \skewchar\fourteensy='60
\font\twelvesy=cmsy10 at 12pt               \skewchar\twelvesy='60
\font\ninesy=cmsy9                          \skewchar\ninesy='60
\font\sixsy=cmsy6                           \skewchar\sixsy='60

\font\seventeenex=cmex10 scaled\magstep3
\font\fourteenex=cmex10 scaled\magstep2
\font\twelveex=cmex10 at 12pt

\font\ninex=cmex10 at 9pt
\font\sevenex=cmex10 at 9pt
\font\sixex=cmex10 at 6pt
\font\fivex=cmex10 at 5pt

\font\seventeensl=cmsl10 scaled\magstep3
\font\fourteensl=cmsl10 scaled\magstep2
\font\twelvesl=cmsl10 scaled\magstep1
\font\ninesl=cmsl10 at 9pt
\font\sevensl=cmsl10 at 7pt
\font\sixsl=cmsl10 at 6pt
\font\fivesl=cmsl10 at 5pt

\font\seventeenit=cmti12 scaled\magstep2
\font\fourteenit=cmti12 scaled\magstep1
\font\twelveit=cmti12

\font\seventeentt=cmtt12 scaled\magstep2
\font\fourteentt=cmtt12 scaled\magstep1
\font\twelvett=cmtt12

\font\seventeencp=cmcsc10 scaled\magstep3
\font\fourteencp=cmcsc10 scaled\magstep2
\font\twelvecp=cmcsc10 scaled\magstep1
\font\tencp=cmcsc10

\newfam\cpfam

\font\seventeenss=cmss17
\font\fourteenss=cmss12 at 14pt
\font\twelvess=cmss12
\font\tenss=cmss10
\font\niness=cmss9

\font\sevenss=cmss8 at 7pt
\font\sixss=cmss8 at 6pt
\font\fivess=cmss8 at 5pt
\newfam\ssfam
\newdimen\b@gheight             \b@gheight=12pt
\newcount\f@ntkey               \f@ntkey=0
\def\f@m{\afterassignment\samef@nt\f@ntkey=}
\def\samef@nt{\fam=\f@ntkey \the\textfont\f@ntkey\relax}
\def\rm{\f@m0 }
\def\mit{\f@m1 }         
\def\cal{\f@m2 }
\def\it{\f@m\itfam}
\def\sl{\f@m\slfam}
\def\bf{\f@m\bffam}
\def\tt{\f@m\ttfam}
\def\ssf{\f@m\ssfam}
\def\caps{\f@m\cpfam}
\def\seventeenpoint{\relax
    \textfont0=\seventeenrm          \scriptfont0=\twelverm
      \scriptscriptfont0=\ninerm
    \textfont1=\seventeeni           \scriptfont1=\twelvei
      \scriptscriptfont1=\ninei
    \textfont2=\seventeensy          \scriptfont2=\twelvesy
      \scriptscriptfont2=\ninesy
    \textfont3=\seventeenex          \scriptfont3=\twelveex
      \scriptscriptfont3=\ninex
    \textfont\itfam=\seventeenit    
    \textfont\slfam=\seventeensl    
      \scriptscriptfont\slfam=\ninesl
    \textfont\bffam=\seventeenbf     \scriptfont\bffam=\twelvebf
      \scriptscriptfont\bffam=\ninebf
    \textfont\ttfam=\seventeentt
    \textfont\cpfam=\seventeencp
    \textfont\ssfam=\seventeenss     \scriptfont\ssfam=\twelvess
      \scriptscriptfont\ssfam=\niness
    \samef@nt
    \b@gheight=17pt
    \setbox\strutbox=\hbox{\vrule height 0.85\b@gheight
                                depth 0.35\b@gheight width\z@ }}
\def\fourteenpoint{\relax
    \textfont0=\fourteencp          \scriptfont0=\tenrm
      \scriptscriptfont0=\sevenrm
    \textfont1=\fourteeni           \scriptfont1=\teni
      \scriptscriptfont1=\seveni
    \textfont2=\fourteensy          \scriptfont2=\tensy
      \scriptscriptfont2=\sevensy
    \textfont3=\fourteenex          \scriptfont3=\twelveex
      \scriptscriptfont3=\tenex
    \textfont\itfam=\fourteenit     \scriptfont\itfam=\tenit
    \textfont\slfam=\fourteensl     \scriptfont\slfam=\tensl
      \scriptscriptfont\slfam=\sevensl
    \textfont\bffam=\fourteenbf     \scriptfont\bffam=\tenbf
      \scriptscriptfont\bffam=\sevenbf
    \textfont\ttfam=\fourteentt
    \textfont\cpfam=\fourteencp
    \textfont\ssfam=\fourteenss     \scriptfont\ssfam=\tenss
      \scriptscriptfont\ssfam=\sevenss
    \samef@nt
    \b@gheight=14pt
    \setbox\strutbox=\hbox{\vrule height 0.85\b@gheight
                                depth 0.35\b@gheight width\z@ }}
\def\twelvepoint{\relax
    \textfont0=\twelverm          \scriptfont0=\ninerm
      \scriptscriptfont0=\sixrm
    \textfont1=\twelvei           \scriptfont1=\ninei
      \scriptscriptfont1=\sixi
    \textfont2=\twelvesy           \scriptfont2=\ninesy
      \scriptscriptfont2=\sixsy
    \textfont3=\twelveex          \scriptfont3=\ninex
      \scriptscriptfont3=\sixex
    \textfont\itfam=\twelveit    
    \textfont\slfam=\twelvesl    
      \scriptscriptfont\slfam=\sixsl
    \textfont\bffam=\twelvebf     \scriptfont\bffam=\ninebf
      \scriptscriptfont\bffam=\sixbf
    \textfont\ttfam=\twelvett
    \textfont\cpfam=\twelvecp
    \textfont\ssfam=\twelvess     \scriptfont\ssfam=\niness
      \scriptscriptfont\ssfam=\sixss
    \samef@nt
    \b@gheight=12pt
    \setbox\strutbox=\hbox{\vrule height 0.85\b@gheight
                                depth 0.35\b@gheight width\z@ }}
\def\tenpoint{\relax
    \textfont0=\tenrm          \scriptfont0=\sevenrm
      \scriptscriptfont0=\fiverm
    \textfont1=\teni           \scriptfont1=\seveni
      \scriptscriptfont1=\fivei
    \textfont2=\tensy          \scriptfont2=\sevensy
      \scriptscriptfont2=\fivesy
    \textfont3=\tenex          \scriptfont3=\sevenex
      \scriptscriptfont3=\fivex
    \textfont\itfam=\tenit     \scriptfont\itfam=\seveni
    \textfont\slfam=\tensl     \scriptfont\slfam=\sevensl
      \scriptscriptfont\slfam=\fivesl
    \textfont\bffam=\tenbf     \scriptfont\bffam=\sevenbf
      \scriptscriptfont\bffam=\fivebf
    \textfont\ttfam=\tentt
    \textfont\cpfam=\tencp
    \textfont\ssfam=\tenss     \scriptfont\ssfam=\sevenss
      \scriptscriptfont\ssfam=\fivess
    \samef@nt
    \b@gheight=10pt
    \setbox\strutbox=\hbox{\vrule height 0.85\b@gheight
                                depth 0.35\b@gheight width\z@ }}
%
%
%
\normalbaselineskip = 15pt plus 0.2pt minus 0.1pt 
\normallineskip = 1.5pt plus 0.1pt minus 0.1pt
\normallineskiplimit = 1.5pt
\newskip\normaldisplayskip
\normaldisplayskip = 15pt plus 5pt minus 10pt 
\newskip\normaldispshortskip
\normaldispshortskip = 6pt plus 5pt
\newskip\normalparskip
\normalparskip = 6pt plus 2pt minus 1pt
\newskip\skipregister
\skipregister = 5pt plus 2pt minus 1.5pt
\newif\ifsingl@    \newif\ifdoubl@
\newif\iftwelv@    \twelv@true
\def\singlespace{\singl@true\doubl@false\spaces@t}
\def\doublespace{\singl@false\doubl@true\spaces@t}
\def\normalspace{\singl@false\doubl@false\spaces@t}
\def\Tenpoint{\tenpoint\twelv@false\spaces@t}
\def\Twelvepoint{\twelvepoint\twelv@true\spaces@t}
\def\spaces@t{\relax
      \iftwelv@ \ifsingl@\subspaces@t3:4;\else\subspaces@t1:1;\fi
       \else \ifsingl@\subspaces@t3:5;\else\subspaces@t4:5;\fi \fi
      \ifdoubl@ \multiply\baselineskip by 5
         \divide\baselineskip by 4 \fi }
\def\subspaces@t#1:#2;{
      \baselineskip = \normalbaselineskip
      \multiply\baselineskip by #1 \divide\baselineskip by #2
      \lineskip = \normallineskip
      \multiply\lineskip by #1 \divide\lineskip by #2
      \lineskiplimit = \normallineskiplimit
      \multiply\lineskiplimit by #1 \divide\lineskiplimit by #2
      \parskip = \normalparskip
      \multiply\parskip by #1 \divide\parskip by #2
      \abovedisplayskip = \normaldisplayskip
      \multiply\abovedisplayskip by #1 \divide\abovedisplayskip by #2
      \belowdisplayskip = \abovedisplayskip
      \abovedisplayshortskip = \normaldispshortskip
      \multiply\abovedisplayshortskip by #1
        \divide\abovedisplayshortskip by #2
      \belowdisplayshortskip = \abovedisplayshortskip
      \advance\belowdisplayshortskip by \belowdisplayskip
      \divide\belowdisplayshortskip by 2
      \smallskipamount = \skipregister
      \multiply\smallskipamount by #1 \divide\smallskipamount by #2
      \medskipamount = \smallskipamount \multiply\medskipamount by 2
      \bigskipamount = \smallskipamount \multiply\bigskipamount by 4 }
\def\normalbaselines{ \baselineskip=\normalbaselineskip
   \lineskip=\normallineskip \lineskiplimit=\normallineskip
   \iftwelv@\else \multiply\baselineskip by 4 \divide\baselineskip by 5
     \multiply\lineskiplimit by 4 \divide\lineskiplimit by 5
     \multiply\lineskip by 4 \divide\lineskip by 5 \fi }
\Twelvepoint  
%
\interlinepenalty=50
\interfootnotelinepenalty=5000
\predisplaypenalty=9000
\postdisplaypenalty=500
\hfuzz=1pt
\vfuzz=0.2pt
\dimen\footins=24 truecm 
\hoffset=10.5truemm 
\voffset=-8.5 truemm 
%
%
%
%
%
%
\def\footnote#1{\edef\@sf{\spacefactor\the\spacefactor}#1\@sf
      \insert\footins\bgroup\singl@true\doubl@false\Tenpoint
      \interlinepenalty=\interfootnotelinepenalty \let\par=\endgraf
        \leftskip=\z@skip \rightskip=\z@skip
        \splittopskip=10pt plus 1pt minus 1pt \floatingpenalty=20000
        \smallskip\item{#1}\bgroup\strut\aftergroup\@foot\let\next}
\skip\footins=\bigskipamount 
\dimen\footins=24truecm 
\newcount\fnotenumber
\def\clearfnotenumber{\fnotenumber=0}
\def\fnote{\advance\fnotenumber by1 \footnote{$^{\the\fnotenumber}$}}
\clearfnotenumber
%
%
\newcount\secnumber
\newcount\appnumber
\newif\ifs@c 
\newif\ifs@cd 
\s@cdtrue 
\def\unsectioned{\s@cdfalse\let\section=\subsection}
\def\clearappnumber{\appnumber=64}
\def\clearsecnumber{\secnumber=0}
\newskip\sectionskip         \sectionskip=\medskipamount
\newskip\headskip            \headskip=8pt plus 3pt minus 3pt
\newdimen\sectionminspace    \sectionminspace=10pc
\newdimen\referenceminspace  \referenceminspace=25pc
\def\Titlestyle#1{\par\begingroup \interlinepenalty=9999
     \leftskip=0.02\hsize plus 0.23\hsize minus 0.02\hsize
     \rightskip=\leftskip \parfillskip=0pt
     \advance\baselineskip by 0.5\baselineskip
     \hyphenpenalty=9000 \exhyphenpenalty=9000
     \tolerance=9999 \pretolerance=9000
     \spaceskip=0.333em \xspaceskip=0.5em
     \seventeenpoint
  \noindent #1\par\endgroup }
\def\titlestyle#1{\par\begingroup \interlinepenalty=9999
     \leftskip=0.02\hsize plus 0.23\hsize minus 0.02\hsize
     \rightskip=\leftskip \parfillskip=0pt
     \hyphenpenalty=9000 \exhyphenpenalty=9000
     \tolerance=9999 \pretolerance=9000
     \spaceskip=0.333em \xspaceskip=0.5em
     \fourteenpoint
   \noindent #1\par\endgroup }
%
\def\spacecheck#1{\dimen@=\pagegoal\advance\dimen@ by -\pagetotal
   \ifdim\dimen@<#1 \ifdim\dimen@>0pt \vfil\break \fi\fi}
\def\section#1{\cleareqnumber \s@ctrue \global\advance\secnumber by1
   \par \ifnum\the\lastpenalty=30000\else
   \penalty-200\vskip\sectionskip \spacecheck\sectionminspace\fi
   \noindent {\caps\enspace\S\the\secnumber\quad #1}\par
   \nobreak\vskip\headskip \penalty 30000 }
\def\subsection#1{\par
   \ifnum\the\lastpenalty=30000\else \penalty-100\smallskip
   \spacecheck\sectionminspace\fi
   \noindent\undertext{#1}\enspace \vadjust{\penalty5000}}

\def\undertext#1{\vtop{\hbox{#1}\kern 1pt \hrule}}
\def\subsubsection#1{\par
   \ifnum\the\lastpenalty=30000\else \penalty-100\smallskip \fi
   \noindent\hbox{#1}\enspace \vadjust{\penalty5000}}

\def\appendix#1{\cleareqnumber \s@cfalse \global\advance\appnumber by1
   \par \ifnum\the\lastpenalty=30000\else
   \penalty-200\vskip\sectionskip \spacecheck\sectionminspace\fi
   \noindent {\caps\enspace Appendix \char\the\appnumber\quad #1}\par
   \nobreak\vskip\headskip \penalty 30000 }
\clearsecnumber
\clearappnumber
%
%
\def\ack{\par\penalty-100\medskip \spacecheck\sectionminspace
   \line{\iftwelv@\fourteencp\else\twelvecp\fi\hfil ACKNOWLEDGEMENTS\hfil}%
\nobreak\vskip\headskip }
\def\refs{\begingroup \par\penalty-100\medskip \spacecheck\sectionminspace
   \line{\iftwelv@\fourteencp\else\twelvecp\fi\hfil REFERENCES\hfil}%
\nobreak\vskip\headskip \frenchspacing }
\def\endrefs{\par\endgroup}
%
\newcount\refnumber
\def\clearrefnumber{\refnumber=0}  \clearrefnumber
\newwrite\R@fs                              
\immediate\openout\R@fs=\jobname.references 
\def\closerefs{\immediate\closeout\R@fs} 
\def\refsout{\closerefs\refs
\catcode`\@=11                          
\input\jobname.references               
\catcode`\@=12			        
\endrefs}
\def\refitem#1{\item{{\bf #1}}}
\def\ifundefined#1{\expandafter\ifx\csname#1\endcsname\relax}
%
%
\def\[#1]{\ifundefined{#1R@FNO}%
\global\advance\refnumber by1%
\expandafter\xdef\csname#1R@FNO\endcsname{[\the\refnumber]}%
\immediate\write\R@fs{\noexpand\refitem{\csname#1R@FNO\endcsname}%
\noexpand\csname#1R@F\endcsname}\fi{\bf \csname#1R@FNO\endcsname}}
\def\refdef[#1]#2{\expandafter\gdef\csname#1R@F\endcsname{{#2}}}
%
%
%
%
%
%
\newcount\eqnumber
\def\cleareqnumber{\eqnumber=0}
\newif\ifal@gn \al@gnfalse  
\def\veqnalign#1{\al@gntrue \vbox{\eqalignno{#1}} \al@gnfalse}
\def\eqnalign#1{\al@gntrue \eqalignno{#1} \al@gnfalse}
\def\(#1){\relax%
\ifundefined{#1@Q}
 \global\advance\eqnumber by1
 \ifs@cd
  \ifs@c
   \expandafter\xdef\csname#1@Q\endcsname{{%
\noexpand\rm(\the\secnumber .\the\eqnumber)}}
  \else
   \expandafter\xdef\csname#1@Q\endcsname{{%
\noexpand\rm(\char\the\appnumber .\the\eqnumber)}}
  \fi
 \else
  \expandafter\xdef\csname#1@Q\endcsname{{\noexpand\rm(\the\eqnumber)}}
 \fi
 \ifal@gn
    & \csname#1@Q\endcsname
 \else
    \eqno \csname#1@Q\endcsname
 \fi
\else%
\csname#1@Q\endcsname\fi\global\let\@Q=\relax}
%
%
\newif\iffrontpage \frontpagefalse
\headline={\hfil}
\footline={\iffrontpage\hfil\else \hss\twelverm
-- \folio\ --\hss \fi }
\def\monthname{\relax\ifcase\month 0/\or January\or February\or
   March\or April\or May\or June\or July\or August\or September\or
   October\or November\or December\else\number\month/\fi}
\hsize=14 truecm
\vsize=22 truecm
\skip\footins=\bigskipamount
\normalspace
%
%
%
\newskip\frontpageskip
\newif\ifp@bblock \p@bblocktrue
\newif\ifm@nth \m@nthtrue
\newtoks\pubnum
\newtoks\pubtype
\newtoks\m@nthn@me
\newcount\Ye@r
\advance\Ye@r by \year
\advance\Ye@r by -1900
\def\Year#1{\Ye@r=#1}
\def\Month#1{\m@nthfalse \m@nthn@me={#1}}
\def\m@nthname{\ifm@nth\monthname\else\the\m@nthn@me\fi}
\def\titlepage{\global\frontpagetrue\hrule height\z@ \relax
               \ifp@bblock\pubblock\fi\relax }
\def\endtitlepage{\vfil\break
                  \frontpagefalse} 
\def\bonntitlepage{\global\frontpagetrue\hrule height\z@ \relax
               \ifp@bblock\pubblock\fi\relax }
\frontpageskip=12pt plus .5fil minus 2pt
\pubtype={\iftwelv@\twelvesl\else\tensl\fi\ (Preliminary Version)}
\pubnum={?}
\def\nopubblock{\p@bblockfalse}
\def\pubblock{\line{\hfil\iftwelv@\twelverm\else\tenrm\fi%
BONN--HE--\number\Ye@r--\the\pubnum\the\pubtype}
              \line{\hfil\iftwelv@\twelverm\else\tenrm\fi%
\m@nthname\ \number\year}}
\def\title#1{\vskip\frontpageskip\Titlestyle{\caps #1}\vskip3\headskip}
\def\author#1{\vskip.5\frontpageskip\titlestyle{\caps #1}\nobreak}

\def\address#1{\par\kern 5pt\titlestyle{
\it #1}}
\def\andaddress{\par\kern 5pt \centerline{\sl and} \address}

\def\KUL{\address{Instituut voor Theoretische Fysica, Universiteit Leuven\break
Celestijnenlaan 200 D, B--3001 Heverlee, BELGIUM}}
\def\Bonn{\address{Physikalisches Institut der Universit\"at Bonn\break
Nu{\ss}allee 12, W--5300 Bonn 1, GERMANY}}
\def\abstract#1{\par\dimen@=\prevdepth \hrule height\z@ \prevdepth=\dimen@
   \vskip\frontpageskip\spacecheck\sectionminspace
   \centerline{\iftwelv@\fourteencp\else\twelvecp\fi ABSTRACT}\vskip\headskip
   {\noindent #1}}
%

%
%
%
\def\leaderfill{\leaders\hbox to 1em{\hss.\hss}\hfill}
\def\boxit#1{\vcenter{\hrule\hbox{\vrule\kern8pt
      \vbox{\kern8pt#1\kern8pt}\kern8pt\vrule}\hrule}}

%
%
%
\def\ref#1{{\bf [#1]}}
\def\ie{{\it i.e.\/}}
\def\nl{\hfil\break}
%
%
%
%
%
\newif\ifm@thstyle \m@thstylefalse
\def\mathstyle{\m@thstyletrue}
\def\proclaim#1#2\par{\smallbreak\begingroup
\advance\baselineskip by -0.25\baselineskip%
\advance\belowdisplayskip by -0.35\belowdisplayskip%
\advance\abovedisplayskip by -0.35\abovedisplayskip%
    \noindent{\caps#1.\enspace}{#2}\par\endgroup%
\smallbreak}
\def\m@kem@th<#1>#2#3{%
\ifm@thstyle \global\advance\eqnumber by1
 \ifs@cd
  \ifs@c
   \expandafter\xdef\csname#1\endcsname{{%
\noexpand #2\ \the\secnumber .\the\eqnumber}}
  \else
   \expandafter\xdef\csname#1\endcsname{{%
\noexpand #2\ \char\the\appnumber .\the\eqnumber}}
  \fi
 \else
  \expandafter\xdef\csname#1\endcsname{{\noexpand #2\ \the\eqnumber}}
 \fi
 \proclaim{\csname#1\endcsname}{#3}
\else
 \proclaim{#2}{#3}
\fi}
%
%
%
%
%
%
\def\Thm<#1>#2{\m@kem@th<#1M@TH>{Theorem}{\sl#2}}
\def\Prop<#1>#2{\m@kem@th<#1M@TH>{Proposition}{\sl#2}}
\def\Def<#1>#2{\m@kem@th<#1M@TH>{Definition}{\rm#2}}
\def\Lem<#1>#2{\m@kem@th<#1M@TH>{Lemma}{\sl#2}}
\def\Cor<#1>#2{\m@kem@th<#1M@TH>{Corollary}{\sl#2}}
\def\Conj<#1>#2{\m@kem@th<#1M@TH>{Conjecture}{\sl#2}}
\def\Rmk<#1>#2{\m@kem@th<#1M@TH>{Remark}{\rm#2}}
\def\Exm<#1>#2{\m@kem@th<#1M@TH>{Example}{\rm#2}}
\def\Qry<#1>#2{\m@kem@th<#1M@TH>{Query}{\it#2}}
%
%
\let\Pf=\Proof

\def\<#1>{\csname#1M@TH\endcsname}
%
%
\def\qed{\vrule width 0.7em height 0.6em depth 0.2em}
\def\QED{\enspace\qed}
\def\lapprox{\hbox{\lower3pt\hbox{$\buildrel<\over\sim$}}}
\def\gapprox{\hbox{\lower3pt\hbox{$\buildrel<\over\sim$}}}
\def\quotient#1#2{#1/\lower0pt\hbox{${#2}$}}

%
%
\def\to{\rightarrow}
%

%
%
%
\def\reals{\mathord{\bf R}} 
%
%
\def\underrightarrow#1{\vtop{\ialign{##\crcr
      $\hfil\displaystyle{#1}\hfil$\crcr
      \noalign{\kern-\p@\nointerlineskip}
      \rightarrowfill\crcr}}} 
\def\underleftarrow#1{\vtop{\ialign{##\crcr
      $\hfil\displaystyle{#1}\hfil$\crcr
      \noalign{\kern-\p@\nointerlineskip}
      \leftarrowfill\crcr}}}  

%
%
%
%
%
\def\der#1#2{{{d #1}\over {d #2}}}
%
%
%
%
%
\newdimen\unit
\newdimen\redunit
%
%
\def\p@int#1:#2 #3 {\rlap{\kern#2\unit
     \raise#3\unit\hbox{#1}}}
%
%
\def\th@r{\vrule height0\unit depth.1\unit width1\unit}
\def\bh@r{\vrule height.1\unit depth0\unit width1\unit}
\def\lv@r{\vrule height1\unit depth0\unit width.1\unit}
\def\rv@r{\vrule height1\unit depth0\unit width.1\unit}
%
%
\def\t@ble@u{\hbox{\p@int\bh@r:0 0
                   \p@int\lv@r:0 0
                   \p@int\rv@r:.9 0
                   \p@int\th@r:0 1
                   }
             }
%
%
\def\t@bleau#1#2{\rlap{\kern#1\redunit
     \raise#2\redunit\t@ble@u}}
%
%
\newcount\n
\newcount\m
\def\makecol#1#2#3{\n=0 \m=#3
  \loop\ifnum\n<#1{}\advance\m by -1 \t@bleau{#2}{\number\m}\advance\n by 1
\repeat}
%
%
\def\makerow#1#2#3{\n=0 \m=#3
 \loop\ifnum\n<#1{}\advance\m by 1 \t@bleau{\number\m}{#2}\advance\n by 1
\repeat}
%
%
\def\checkunits{\ifinner \unit=6pt \else \unit=8pt \fi
                \redunit=0.9\unit } 
\def\ytsym#1{\checkunits\kern-.5\unit
  \vcenter{\hbox{\makerow{#1}{0}{0}\kern#1\unit}}\kern.5em} 
\def\ytant#1{\checkunits\kern.5em
  \vcenter{\hbox{\makecol{#1}{0}{0}\kern1\unit}}\kern.5em} 
\def\ytwo#1#2{\checkunits
  \vcenter{\hbox{\makecol{#1}{0}{0}\makecol{#2}{1}{0}\kern2\unit}}
                  \ } 
\def\ythree#1#2#3{\checkunits
  \vcenter{\hbox{\makecol{#1}{0}{0}\makecol{#2}{1}{0}\makecol{#3}{2}{0}%
\kern3\unit}}
                  \ } 
%
%
%

\def\NPB#1#2#3{{\sl Nucl. Phys.} {\bf B#1} (#2) #3}

\def\CMP#1#2#3{{\sl Comm. Math. Phys.} {\bf #1} (#2) #3}

\def\PLB#1#2#3{{\sl Phys. Lett.} {\bf #1B} (#2) #3}

\def\FAaIA#1#2#3{{\sl Functional Analysis and Its Application} {\bf #1} (#2)
#3}

\def\Invm#1#2#3{{\sl Invent. math.} {\bf #1} (#2) #3}

\def\IJMPA#1#2#3{{\sl Int. J. Mod. Phys.} {\bf A#1} (#2) #3}

\def\TMP#1#2#3{{\sl Theor. Mat. Phys.} {\bf #1} (#2) #3}

\def\JSM#1#2#3{{\sl J. Soviet Math.} {\bf #1} (#2) #3}

\catcode`\@=12 
%
%
%
%
\def\W{\mathord{\ssf W}}
\def\w{\mathord{\ssf w}}
\def\wB{\mathord{\ssf wB}}
\def\wC{\mathord{\ssf wC}}
\def\gd{\mathord{\ssf gd}}
\def\diffs#1{\mathord{\ssf diff}(S^{#1})}
\def\d{\partial}
\let\pb=\anticomm

\def\fr#1/#2{\hbox{${#1}\over{#2}$}}
\def\mod{\mathop{\rm mod}}
\refdef[Adler]{M.~Adler, \Invm{50}{1981}{403}.}
\refdef[Dickey]{L.~A.~Dickey, {\sl Integrable equations and Hamiltonian
systems}, World Scientific Publ.~Co.}
\refdef[GD]{I.~M.~Gel'fand and L.~A.~Dickey, {\sl A family of Hamiltonian
structures connected with integrable nonlinear differential equations},
Preprint 136, IPM AN SSSR, Moscow (1978).}
\refdef[DS]{V.~G.~Drinfel'd and V.~V.~Sokolov, \JSM{30}{1984}{1975}.}
\refdef[FL]{V.~A.~Fateev and S.~L.~Lukyanov, \IJMPA{3}{1988}{507}.}
\refdef[ClassLim]{J. M. Figueroa-O'Farrill and E. Ramos,
\PLB{282}{1992}{357} ({\tt hep-th/9202040}.}
\refdef[Zam]{A. B. Zamolodchikov, \TMP{65}{1986}{1205}.}
\refdef[Class]{J.~M.~Figueroa-O'Farrill and E.~Ramos, {\sl
Classical $\W$-algebras from dispersionless Lax hierarchies},
Preprint-KUL-TF-92/6.}
\refdef[Krich]{I. Krichever, \CMP{143}{1992}{415}.}
\refdef[WGeom]{C.M. Hull, \PLB{269}{1991}{257};\nl
G.~Sotkov and M.~Stanishkov, \NPB{356}{1991}{439};\nl
G.~Sotkov, M.~Stanishkov and C.~J.~Zhu, \NPB{356}{1991}{245};\nl
J.~Gervais \& Y.~Matsuo, \PLB{282}{1992}{309} ({\tt hep-th/9110028});
Preprint LPTENS-91/35, NBI-HE-91-50 ({\tt hep-th/9201026});\nl
J.~de~Boer and J.~Goeree, THU--92/14 ({\tt hep-th/9206098}).}
\refdef[Radul]{A.~O.~Radul, \FAaIA{25}{1991}{25}.}
\overfullrule=0pt
\unsectioned
\def\pubblock{ \line{\hfil\twelverm BONN--HE--92--27}
               \line{\hfil\twelverm KUL--TF--92/34}
               \line{\hfil\twelvett hep-th/9209002}
               \line{\hfil\twelverm August 1992}}
\titlepage
\title{A Geometrical Interpretation of Classical $\W$-Transformations}
\vskip 1.cm
\author{Jos\'e~M.~Figueroa-O'Farrill\footnote{$^\natural$}{\tt
e-mail: figueroa@pib1.physik.uni-bonn.de\hfil},
Sonia~Stanciu\footnote{$^\flat$} {\tt e-mail:
stanciu@pib1.physik.uni-bonn.de\hfil}}
\Bonn
\vskip 0.5cm
\centerline{\it and}
\author{Eduardo Ramos\footnote{$^\sharp$}{{\tt e-mail:
fgbda06@blekul11.bitnet\hfil}\break {\rm (Address after October 1992:
Queen Mary and Westfield College, UK)\hfil}}}
\KUL

\abstract{We give a simple geometrical interpretation of
classical $\W$-transformations as deformations of constant energy
surfaces by canonical transformations on a two-dimensional phase
space.}

\endtitlepage

\section{Introduction}

The purpose of this letter is to give a simple and readable account of
the geometric significance of classical $\W$-transformations\fnote{For
different approaches to this problem see \[WGeom].} ($\w$-morphisms).
We should first explain what is meant by a $\w$-morphism from the
strictly algebraic point of view.

$\W$-algebras were first introduced in \[Zam], where it is shown,
using the bootstrap method, that the extension of the Virasoro algebra
by a field of spin 3 ($W$) yielded a non-linear associative algebra,
denoted since then by $\W_3$.  It is well-known that
$$Q_{\epsilon}=\oint dz \,\epsilon (z)T(z)\(diff)$$
is the generator of conformal transformations (or, equivalently, of
diffeomorphisms of the circle). But, what is the geometrical
significance, if any, of the transformation generated by
$$Q_{\eta}=\oint dz\,\eta(z)W(z)~?\(Wtransf)$$

Soon after Zamolodchikov's paper, Fateev and Lukyanov \[FL]
recognized that the second hamiltonian structure of the Boussinesque
hierarchy (the Gel'fand--Dickey algebra associated to the Boussinesque
operator) is a classical realisation of the $\W_3$-algebra.  Then
they were able, using the formalism of Drinfel'd and
Sokolov \[DS], to generalise the results of Zamolodchikov to
construct $\W_n$-algebras; \ie, extended conformal algebras with
fields of integer spins from 3 to $n$. Of course, from the geometrical
point of view this development was not of much help: it only seemed to
complicate matters even further opening the question of what is the
geometrical meaning of all these new $\W_n$-transformations.
Nevertheless, it was a crucial development from the point of view of
the algebraic theory, putting at our disposal all the powerful
machinery of integrable systems of the KdV-type. (For a comprehensive
review see \[Dickey].)

Recently, we became interested in the problem of looking for simpler
algebraic structures that would still retain the essential features of
$\W$-algebras.  In \[ClassLim] it was proven that it is possible to
define a classical limit of the Gel'fand--Dickey algebras and their
reductions; thus providing a natural simplification.  These are
nonlinear extensions of $\diffs1$ by tensors $\{u_j\}$ of weights
$3,4,\cdots,n$.  The name classical $\W$-algebras is justified in the
sense that upon quantisation the full structure of the $\W$-algebra is
recovered.  We would also like to stress that these classical
$\W$-algebras are of great interest in their own right.  For example,
they play a fundamental r\^ole in the context of planar $2{-}D$
gravity \[Class], as well as in $2{-}D$ topological field
theory \[Krich].

The simplifications introduced by the classical limit notwithstanding,
it remains to elucidate the geometrical meaning of the $\w$-morphisms
generated, under Poisson bracket, by
$$Q^j_{\epsilon}=\int dx\,\epsilon (x)u_j(x)~.\(wmorph)$$
We will see in what follows that $\w$-morphisms can be interpreted as
deformations of constant-energy surfaces in a two-dimensional
phase-space induced by infinitesimal canonical transformations.

But before getting into the details we would like to remark that this
result has been inspired by the relation found by Radul \[Radul]
between the algebra of differential operators on the circle and the
Gel'fand--Dickey algebras; although our presentation will not make
this explicit.

\section{Geometric Setup}

Consider a two-dimensional phase space $M$ and a smooth function $H$
on $M$, which to fix ideas we can think of as a hamiltonian.  Let
$\lambda$ be a (regular) value of $H$, so that the constant energy
(one-dimensional) surface $Z=H^{-1}(\lambda)$ is a submanifold of $M$.
Let $L\equiv H-\lambda$.  Then $Z$ is the zero locus of $L$ and $L$
generates the ideal ${\cal I}_Z$ of functions vanishing on $Z$.  This
ideal consists of functions $FL$, where $F$ is any smooth function on
$M$.  It is clear that such functions vanish on $Z$ and it can be
proven that these are all the functions which do.  Any function on $Z$
extends to a function on all of $M$ and the difference of any two such
extensions is a function vanishing on $Z$.  In other words, there is a
one-to-one correspondence between the functions ${\cal F}(Z)$ on $Z$
and the quotient ${\cal F}(M)/{\cal I}_Z$.  We let $\pi: {\cal F}(M)
\to {\cal F}(M)/{\cal I}_Z$ denote the map which sends a function on
$M$ to its equivalence class modulo ${\cal I}_Z$.  In the next
section, and for the class of functions we shall consider, we exhibit
an explicit model for this quotient.

We now investigate the effect of canonical transformations
(symplectomorphisms) on the constant energy surface $Z$.
We can analyze deformations of $Z$ by looking at how the function $L$
behaves on $Z$ under symplectomorphisms.

Infinitesimal symplectomorphisms are locally generated by functions on
$M$.  In fact, given a function $S$ on $M$, it gives rise to a vector
field $\delta_S$ defined such that acting on a function $F$,
$$\delta_S F = \pb{S}{F}~.\()$$
If $S$ vanishes on $Z$, then $\delta_S$ is tangent to $Z$.  In fact,
such an $S$ can be written as $GL$ and hence
$$\delta_S L = \pb{GL}{L} = \pb{G}{L}L~,\()$$
which vanishes on $Z$.  (Physically this is nothing but energy
conservation.)  Therefore infinitesimal symplectomorphisms generated
by functions in ${\cal I}_Z$ do not change $Z$.  In other words,
nontrivial deformations of $Z$ induced from symplectomorphisms are
locally generated by ${\cal F}(M)/{\cal I}_Z$.
Therefore, on $Z$, the function $L$ transforms as
$$\delta_S L \equiv \pi(\pb{\pi(S)}{L})~.\(gendef)$$
For a specific choice of hamiltonian, we will now see that \(gendef)
defines $\w$-morphisms associated to the classical $\W$-algebras:
$\gd_n$ and its reduction $\w_n$.

\section{Classical $\W$-Transformations}

To fix the ideas, we now specialize to $M = S^1\times \reals$ a
cylinder.\fnote{This represents no loss of generality.  The only other
connected two-dimensional phase space (\ie, cotangent bundle) is the
plane, and we can recover this case by simply working locally on the
cylinder.}  In other words, $M$ is the phase space whose configuration
space is a circle.  A coordinate system $q$ for the circle gives rise
to a coordinate system $(q,p)$ for $M$ in such a way that
$\pb{p}{q}=1$.  Moreover any other coordinate system $Q$ on the circle
is related to $q$ by a diffeomorphism, and the associated
coordinate system $(Q,P)$ is related to $(q,p)$ by a canonical
transformation.  Explicitly, if $q \mapsto Q(q)$, then $p\mapsto P =
p/Q'$, where $Q' = \der{Q}{q}$.  This preserves the fundamental
one-form $\theta = pdq = PdQ$ and hence the symplectic form $\omega =
d\theta$ whence the Poisson brackets.

As our function $L$ we choose one of the form $L(q,p) = p^n +
\sum_{i=1}^n u_i(q) p^{n-i}$, where $u_i$ are arbitrary functions.
Under a change of coordinates $(q,p) \to (Q,P)$,
$$L(q,p) = (Q')^n\left(P^n + \sum_{i=1}^n U_i(Q)
P^{n-i}\right)~,\(newL)$$
where $U_i$ and $u_i$ are related by
$$u_i(q) = (Q')^i U_i(Q)~.\(utransf)$$
Since $q\mapsto Q(q)$ is a diffeomorphism, $Q'$ is nowhere vanishing,
hence the submanifold $Z$ which is defined as the zero locus of $L$ in
the coordinates $(q,p)$ is defined, in the coordinates $(Q,P)$, as the
zero locus of the function $P^n + \sum_{i=1}^n U_i(Q) P^{n-i}$, which
has the same form.  Thus these constant-energy surfaces have an
invariant geometric meaning.

In order to have an algebraic handle on the situation, we will work
with functions whose dependence on $p$ is polynomial.  Under a change
of coordinates $(q,p) \to (Q,P)$, polynomials in $p$ go over to
polynomials in $P$.  Let ${\cal E}$ denote the subring of these
functions.  Notice that $L$ belongs to ${\cal E}$.  We let ${\cal
J}_Z$ denote the ideal of ${\cal E}$ generated by $L$.  Since $p^n = L
- \sum_{i=1}^n u_i(q) p^{n-i}$, we notice that modulo ${\cal J}_Z$ we
can always reduce any function in ${\cal E}$ to one with at most $n-1$
powers of $p$.  In other words, ${\cal E}/{\cal J}_Z$ is in one-to-one
correspondence with the functions of the form $\sum_{i=0}^{n-1}f_i(q)
p^i$.  We now give an explicit expression for this representative.
For this we will have to introduce $L^{-1}_{(r)}$---polynomial
functions in $p$ and $p^{-1}$ and which correspond to finite
truncations of the formal inverse of $L$.  Explicitly,
$$L^{-1} = p^{-n}\sum_{k\geq 0}^{\infty}(-1)^k \left(\sum_{j=0}^{n-1}
u_j(q)p^{j-n}\right)^k~,\(linv)$$
and $L_{(r)}^{-1}$ is defined by $L^{-1}= L_{(r)}^{-1} \mod
O(p^{-n-r-1})$ with $r\geq 0$, such that
$$L^{-1}_{(r)}\cdot L= 1 +O(p^{-r-1}).\()$$
Given any polynomial function $F$ in $p$ and $p^{-1}$ we denote by
$F_+$ the part polynomial in $p$ and $F_- = F - F_+$.

\Prop<projection>{Any element $R$ of ${\cal E}$ of order $r$ is
equivalent modulo ${\cal J}_Z$ to a unique polynomial of order at most
$n-1$ given by
$$\pi_L(R) = R-(RL^{-1}_{(r)})_+L
=\left((RL_{(r)}^{-1})_-L\right)_+~.\(projector)$$}

\Pf It is obvious that $\pi_L(R)$ is polynomial in $p$ of order
smaller than $n$ and, moreover, $\pi_L(R) - R \in {\cal J}_Z$.
Uniqueness follows because the order of any function in ${\cal J}_Z$
is equal or bigger than $n$.\QED

This provides us with a concrete model for the equivalence space
${\cal E}/{\cal J}_Z$---namely the space ${\cal E}_{<n}$ of functions
polynomial in $p$ with order strictly less than $n$.  In the sequel,
and in order not to clutter the notation, we will write $L^{-1}$ in
\(projector) to mean the appropriate truncation $L^{-1}_{(r)}$.

We now have at our disposal all the ingredients to establish the link
between the algebraic $\w$-morphisms alluded to in the introduction
and the deformation of constant-energy surfaces.   To this effect, we
compute \(gendef) in this concrete example, where we now make use of
our explicit projector $\pi_L$ instead of $\pi$.  Since
$$\pi_L(S) = ((SL^{-1})_-L)_+~,\()$$
it is natural to reparametrize $\w$-morphisms by
$$X = (SL^{-1})_- \mod p^{-n-1}~,\(reparam)$$
with $\pi_L(S) = (XL)_+$.  We can then write \(gendef) as follows
$$\veqnalign{\delta_X L \equiv \delta_S L &= \pb{(XL)_+}{L} -
(\pb{(XL)_+}{L}L^{-1})_+L\cr
&= \pb{(XL)_+}{L} - \pb{(XL)_+L^{-1}}{L}_+L\cr
&= \pb{(XL)_+}{L} - \pb{X}{L}_+L~,\(cladler)\cr}$$
which as shown in \[ClassLim] is the classical limit of the Adler
map or, equivalently, of the Gel'fand--Dickey brackets---namely
$\gd_n$.  This establishes the equivalence between the algebraic
and geometric approaches to $\w$-morphisms.

In order to obtain now the classical limit $\w_n$ of the $\W_n$
algebras, we need to restrict ourselves to functions $L$ of the form
$$L(q,p) = p^n + \sum_{i=2}^n u_i(q) p^{n-i}~.\(redL)$$
One can always achieve this by a symplectomorphism of the form
$$\eqalign{p &\mapsto p -{1\over n} u_1(q)\cr
q &\mapsto q~,\cr}\()$$
which puts the coefficient of $p^{n-1}$ to zero.  Notice moreover
that, under coordinate changes induced from diffeomorphisms of the
circle, this form of $L$ is preserved.  It then follows that if we
restrict ourselves to infinitesimal symplectomorphisms which preserve
the constraint, \(cladler) define $\w$-morphisms associated with
$\w_n$ \[ClassLim].

Finally, if we restrict to functions $L$ which are odd or even under
the transformation $p\mapsto -p$, and we again only consider
symplectomorphism preserving such property, \(cladler) will induce
$\w$-morphisms associated with the $\wB$ or $\wC$ series,
respectively.

\section{A Simple Example: $\w_3$.}

Consider now, as an example, the function
$$L(q,p) = p^3 + T(q) p + W(q)~.\(wiiiL)$$
The associated classical $\W$-algebra is the $\w_3$-algebra:
$$\veqnalign{\pb{T(x)}{T(y)}^{c\ell} &= - \left[ 2 T(x)\d +
T'(x)\right] \cdot \delta(x-y)~,\cr
\pb{W(x)}{T(y)}^{c\ell} &= - \left[ 3 W(x)\d + W'(x)\right] \cdot
\delta(x-y)~,\()\cr
\noalign{\hbox{and}}
\pb{W(x)}{W(y)}^{c\ell} &= \left[ {2\over3} T(x)\d T(x)\right]
\cdot \delta(x-y)~,\cr}$$
which corrects a typographical error in \[ClassLim].

The algebraic $\w$-morphisms generated by $T$ and $W$ under the above
algebra are given by the usual formulas
$$\veqnalign{\delta_\epsilon^{(T)} F(y) &= \int dx\, \epsilon(x)
\pb{T(x)}{F(y)}^{c\ell}~,\()\cr
\noalign{\hbox{and}}
\delta_\alpha^{(W)} F(y) &= \int dx\, \alpha(x)
\pb{W(x)}{F(y)}^{c\ell}~.\()\cr}$$
With them we can compute the effect of $\w$-morphisms on the
generators themselves.  We obtain
$$\eqalign{\delta_\epsilon^{(T)} T &= 2T\epsilon' + T'\epsilon\,\cr
\delta_\epsilon^{(T)} W &= 3W\epsilon' + W' \epsilon\,\cr
\delta_\alpha^{(W)} T &= 2W'\alpha + 3W\alpha'\,\cr
\delta_\alpha^{(W)} W &= -\fr2/3 (\alpha T)' T~.\cr}\(algwmor)$$

We now compute the deformation of the constant-energy surface $Z$
defined by $L$ using the geometric procedure introduced earlier.  The
most general infinitesimal symplectomorphism which yields a nontrivial
deformation of $Z$ is generated by functions of the form
$$\pi_L(S) = \alpha p^2 + \epsilon p + \beta~.\()$$
Demanding that the symplectomorphism preserve the form \(wiiiL) of L
requires that $\beta = \fr2/3 \alpha T$.  We can now compute \(gendef)
yielding
$$\delta_S L = \left(\delta_\epsilon^{(T)} T + \delta_\alpha^{(W)}
T\right) p + \delta_\epsilon^{(T)} W + \delta_\alpha^{(W)} W~,\()$$
with the variations given by \(algwmor).

\ack

We would like to thank S. Schrans for a careful reading of a previous
draft of the \TeX script. JMF and SS take great pleasure in expressing
their thanks to the Instituut voor Theoretische Fysica of the
Universiteit Leuven for its hospitality and support during the start
of this collaboration.

\refsout
\bye